\documentclass[aps,pra,twocolumn]{revtex4-1} 

\usepackage{verbatim}
\usepackage{dsfont} 
\usepackage{amsmath}
\usepackage{amsthm} 
\usepackage{graphicx}
\usepackage{mathtools} 
\usepackage{bm} 
\usepackage{amssymb} 
\usepackage{hyperref} 
\usepackage{appendix} 
\usepackage[capitalize,nameinlink]{cleveref}
\hypersetup{
    colorlinks=true,
    linkcolor=blue,
    citecolor=blue,
    unicode=true,          
}

\usepackage{blindtext}

\newtheorem{result}{Result}

\newtheorem*{lemma}{Lemma}

\newtheorem*{conjecture}{Conjecture}

\newcommand{\ket}[1]{\left| #1 \right\rangle}

\newcommand{\braket}[1]{\left\langle #1 \right\rangle}
\newcommand{\ketbra}[2]{\left|#1 \rangle \langle #2 \right|}
\DeclareMathOperator{\tr}{Tr}
\usepackage{textpos}

\begin{document}

\title{Locally inaccessible hidden quantum correlations}

\author{Andr\'es F. Ducuara$^{1,2}$} 
\email[]{andres.ducuara@yukawa.kyoto-u.ac.jp}

\author{Cristian E. Susa$^{3}$}
\email[]{cristiansusa@correo.unicordoba.edu.co}

\author{Paul Skrzypczyk$^{4,5}$}
\email[]{paul.skrzypczyk@bristol.ac.uk}

\affiliation{$^{1}$Yukawa Institute for Theoretical Physics, Kyoto University, Kitashirakawa Oiwakecho, Sakyo-ku, Kyoto 606-8502, Japan
\looseness=-1}

\affiliation{$^{2}$Center for Gravitational Physics and Quantum Information, Yukawa Institute for Theoretical Physics, Kyoto University
\looseness=-1} 

\address{$^{3}$Department of Physics and Electronics, University of C\'ordoba, 230002 Monter\'ia, Colombia\looseness=-1}

\affiliation{$^{4}$H.H. Wills Physics Laboratory, University of Bristol, Tyndall Avenue, Bristol, BS8 1TL, United Kingdom 
\looseness=-1}

\affiliation{$^{5}$CIFAR Azrieli Global Scholars program, CIFAR, Toronto, Canada\looseness=-1}

\date{\today}

\begin{abstract}
We prove, modulo a conjecture on quantum steering ellipsoids being true, the existence of the phenomenon of \emph{locally inaccessible hidden quantum correlations}. That is, the existence of two-particle states whose hidden quantum correlations \emph{cannot} be revealed by local filters implemented exclusively on one side of the experiment, but that can still be revealed when \emph{both} parties cooperate in applying judiciously chosen local filters. The quantum correlations here considered are the violation of the CHSH-inequality for Bell-nonlocality and the violation of the $\rm F_3$-inequality for EPR-steering. Specifically, we provide a necessary criterion for guaranteeing the presence of such phenomenon for arbitrary two-qubit states. This criterion in turn relies on the conjecture that the maximal violation of CHSH-inequality and $\rm F_3$-inequality are both upper bounded by functions that depend on the magnitude of the quantum steering ellipsoid  centre. This latter conjecture, although currently lacking an analytical proof, is supported by numerical results. We use this necessary criterion to explicitly show examples of two-qubit states with locally inaccessible hidden quantum correlations and, furthermore, two-qubit states with locally inaccessible \emph{maximal} hidden quantum correlations. 
\end{abstract} 
\maketitle

\begin{textblock*}{3cm}(17cm,-10.5cm)
  \footnotesize YITP-23-83
\end{textblock*}
\vspace{-1cm}
\section{Introduction}

The identification and characterisation of quantum correlations \cite{GA2016} like Bell-nonlocality \cite{QNL2014, book_nonlocality} and EPR-steering \cite{QS2017, review_steering2} is a crucial endeavour for the development of quantum technologies like device-independent and semi-device-independent information protocols \cite{VS2015}. Quantum states which cannot directly produce these type of correlations can still however exhibit \emph{hidden} versions of them, which can be revealed by implementing local filtering operations \cite{SP1995, Gisin1996, FV2001a}. This phenomenon of revealing or extracting quantum correlations by means of local filtering operations, first shown to be possible in the works of Popescu \cite{SP1995} and Gisin \cite{Gisin1996}, has been addressed for revealing various quantum properties such as hidden Bell-nonlocality \cite{SP1995, Gisin1996}, hidden EPR-steering \cite{EXP3}, hidden usefulness for teleportation \cite{FV2001a, LF13}, and for maximally extracting entanglement \cite{LF5}, and have been subject of extensive study \cite{LF3, LF4, LF5, LF6, LF7, LF8, LF9, LF10, LF11, LF12}. Local filtering operations have been successfully implemented and verified in various experimental setups \cite{EXP0, EXP1, EXP2, EXP3, EXP4, EXP5, EXP6}. 

Revealing hidden quantum correlations, like hidden Bell-nonlocality for instance, refers to the situation in which experimentalists Alice and Bob share a bipartite quantum state, which itself cannot be used to violate any Bell-inequality, but that can still be transformed (by means of local filtering operations) into a state which can now violate a Bell-inequality. It happens sometimes however that local filters on either Alice's or Bob's side \emph{alone}, are enough to reveal these hidden correlations. This observation then leads one to wonder about the existence of states for which local filters on either Alice's or Bob's side \emph{alone} are \emph{never} enough for the procedure to work, but that can still however be made to work when \emph{both} parties cooperate in applying local filters. We refer to this latter phenomenon as \emph{locally inaccessible hidden quantum correlations}. The main question explored in this manuscript is whether this type of correlations can actually exist and how to detect them if that is the case. 

Before delving into the ways one can propose to tackle this problem, let us first address some physically motivated scenarios in which these one-sided variants of the standard locally filtered Bell-test can naturally emerge. First, these one-sided variants can be found useful when the parties do not trust each other, and then they each require to be completely certain that the other party did their job in regards to actually applying a local filter. In other words, if Alice and Bob share a state with locally inaccessible hidden correlations, a locally filtered Bell-test displaying a violation of a Bell-inequality will guarantee that local filters were indeed implemented on \emph{both} sides of the experiment (since the state does not allow a violation if only one of the parties acted), and so Alice can be reassured that Bob did his job in applying the required filter and viceversa. Second, from a conceptual point of view, we note that whilst (hidden) Bell-nonlocality is usually regarded as a ``symmetric" property, in the sense that a bipartite state, as a whole, is either (hidden) non-local or not, the existence of quantum states displaying locally inaccessible hidden quantum correlations suggests that quantum states can still posses an intrinsic asymmetry with respect to their nonlocal behaviour. This, since these states would display a hidden nonlocal behaviour when manipulated from Alice's point of view, but not from Bob's. Third and finally, we can similarly explore hidden quantum correlations being \emph{locally accessible}. We can imagine that local filters might actually be infeasible to be implemented by one of the involved parties, say Bob, due to experimental limitations for instance and, therefore, it becomes desirable to consider a scenario addressing the hidden quantum correlations that are locally accessible from Alice's point of view. 

One key difficulty in approaching the above question is to show that a state cannot display a certain type of quantum correlation after \emph{any} local filter has been applied on one side \cite{hirsch_ewhn}. For example, considering Bell-nonlocality, this would requires methods for finding a local-hidden-variable (LHV) model for a state, after any filtering has been performed on one side. Finding LHV models is a notoriously difficult task, even in the absence of filters \cite{PS2016, NB2016, NB2018}, and the only general techniques known do not seem to generalise readily to the case of arbitrary local filters \cite{hirsch_ewhn}. 

In order to gain a first insight into this problem, we consider more specialised scenarios. In particular, the most important Bell-inequality is arguably the Clauser-Horne-Shimony-Holt (CHSH) inequality \cite{CHSH1969}. This is the most basic of all Bell inequalities, which has been the subject of extensive study over the years, and which now also plays a key role in numerous applications in the domain of device-independent (DI) quantum information \cite{VS2015}. Our central point here is thus primarily to explore locally inaccessible hidden violations of the CHSH-inequality. Explicitly, we focus on quantum states which do not violate the CHSH-inequality, and explore whether they can violate it after appropriate one-sided local filters. Such a phenomenon is relevant, for example, to any DI protocol built upon violations of the CHSH-inequality \cite{VS2015}. 

We also apply the same reasoning to the case of EPR-steering. Showing that a state is unsteerable, i.e., that it has a local-hidden-state (LHS) model is again a difficult problem. We thus focus on the simplest fixed steering inequality, the $F_3$-inequality, where Bob measures the three Pauli operators on his qubit system. This inequality is again important from the perspective of one-sided device-independent (1SDI) quantum information, and it is therefore relevant to ask whether there are states which can have locally inaccessible hidden violations of this inequality, meaning that we concern on states which do not violate the $F_3$-inequality, but that can do so after local filtering operations.

We obtain preliminary positive results in this direction by deriving a necessary criterion for witnessing states with locally inaccessible hidden quantum correlations, specifically for the violation of the CHSH-inequality and the $F_3$-inequality. This criterion relies on the validity of a conjecture on quantum steering ellipsoids, first described in the work of Milne {\it et. al.} \cite{AM2014pra}, which is supported by numerical results \cite{AM2014pra}. We use this criterion to provide examples of two-qubit states displaying the phenomena of B-inaccessible hidden correlations, AB-inaccessible hidden correlations, as well as AB-inaccessible \emph{maximal} hidden correlations.

This document is organised as follows. We start in section II with some preliminaries, notation, and the measures of quantum correlations that we are interested in; the violation of the CHSH-inequality for Bell-nonlocality and the violation of the $\rm F_3$-inequality for EPR-steering. This is followed by a succinct description of QSEs. In section III we address standard hidden quantum correlations and formalise what we mean by locally accessible and inaccessible hidden quantum correlations. We then derive our main result in the form of a necessary criterion for detecting locally inaccessible hidden quantum correlations for arbitrary two-qubit states. In section IV we use this criterion to explicitly show examples of states with various such properties. We end up with some conclusions, open problems, and perspectives.

\section{Preliminaries}

We focus on the scenario where experimentalists Alice an Bob share an arbitrary two-qubit state $\rho \in D(\mathds{C}^2 \otimes \mathds{C}^2)$ which can be written as $\rho = \frac{1}{4} \sum _{i,j=0}^3 R_{ij}\sigma_i \otimes \sigma_j$, where $R_{ij} = \tr \left[ (\sigma_i \otimes \sigma_j)\rho \right]$ with $\sigma_0 = \mathds{1}$ and $\{\sigma_i\}, i=1,2,3$ the Pauli matrices. It is convenient to write this real matrix as:
\begin{align}
    R
    =
    \begin{pmatrix} 
    1& \bm{b} ^{\,T}
    \\
    \bm{a}& T
\end{pmatrix},
\label{eq:Rmatrix}
\end{align}
where $\bm{a}=[a_i]$ with $a_i= \tr [(\sigma_i \otimes  \mathds{1})\rho]$, $\bm{b}=[b_i]$ with $b_i= \tr [(\mathds{1}\otimes \sigma_i)\rho]$ are the Bloch vectors of the reduced states and $T_{ij}={\rm{Tr}}[\rho (\sigma_i  \otimes \sigma_j)]$ with $i,j=1,2,3$ the correlation matrix. Bob's reduced state is given by $\rho_B={\rm Tr_A}[\rho]=\frac{1}{2} \left( \mathds{1}+\bm{b} \cdot \bm{\sigma} \right)$ with $\bm \sigma = [\sigma_i]$ and similarly for Alice's. We are interested in the CHSH inequality, which accounts for Bell-nonlocality and the $\rm F_3$-inequality, which accounts for EPR-steering, these two inequalities read:
{\small \begin{align*}
    {\rm M}
    \left(
    \rho, \{A^i,B^j\}
    \right)
    =
    &\frac{1}{2}  (\braket{A^1B^1}+\braket{A^1B^2}+\\
    &+\braket{A^2B^1}-\braket{A^2B^2}) \leq 1,\\
    {\rm F_3}
    \left(
    \rho,\{A^i\}
    \right)
    =&
    \frac{1}{\sqrt{3}}
    \left(
    \braket{A^1\sigma_1}
    +
    \braket{A^2\sigma_2}
    +
    \braket{A^3\sigma_3}
    \right)\leq 1,
\end{align*}}
\noindent 
with the expectation values $\braket{A^iB^j} = \tr[(A^i\otimes B^j)\rho]$ and observables $A^i = \bm{\sigma}\cdot \boldsymbol{\alpha}^i$, $B^j = \bm{\sigma}\cdot \boldsymbol{\beta}^j$, $\boldsymbol{\alpha}^i$ and $\boldsymbol{\beta}^j$ real unit vectors. We want to maximise these functions over all possible measurements, so as to maximally violate these inequalities. The Horodecki criterion \cite{HHH1995} and the Costa-Angelo criterion \cite{CA2016} solve these optimisation problems as:
\begin{align}
    {\rm B}(\rho)
    &=
    \underset{\{A^i,B^j\}}{\rm max} {\rm M}\left(\rho, \{A^i,B^j\}\right)=\sqrt{s_1^2+s_2^2}, \label{eq:HC}\\
    {\rm F_3}(\rho)
    &=
    \underset{\{A^i\}}{\rm max} \,{\rm F_3}\left(\rho, \{A^i\}\right)=\sqrt{s_1^2+s_2^2+s_3^2}, \label{eq:CAC}
\end{align}
 where $\{s_i\}$ are the singular values in decreasing order of the correlation matrix $T$ \eqref{eq:Rmatrix}. Let us now consider that Alice performs a general POVM measurement with $O_A$ outcomes as $E=\{E_e\}$, $e=1,...,O_A$. The POVM effects can be written as $E_e=\frac{1}{2} \left( \mathds{1} + \boldsymbol{\gamma}_e \cdot \bm{\sigma}\right)$ with $|\boldsymbol{\gamma}_e|\leq 1$, $|\boldsymbol{\gamma}_e|$=1 for projective measurements. After the implementation of a particular POVM effect, Bob's reduced state is ``steered" to a state of the form:
\begin{align*}
  \rho^e_B
  =
  \frac{1}{2} 
  \left(
  \mathds{1}
  +
  \bm{b} 
  (
  \boldsymbol{\gamma}_e
  )
  \cdot
  \bm{\sigma} 
  \right), \hspace{0.5cm} 
  \bm{b}
  (
    \boldsymbol{\gamma}_e
  )
  =
  \frac{1}{2p_e} 
  \left(
  \bm{b}
  +
  T^T
  \boldsymbol{\gamma}_e 
  \right),
\end{align*}
with probabilities $p_e = \tr[ ( E_e \otimes \mathds{1}) \rho ]=\frac{1}{2}( 1 + \bm{a} \cdot \boldsymbol{\gamma}_e)$. The Bloch vectors $\bm{b}(\boldsymbol{\gamma}_e)$ turn out to lie on the surface of an ellipsoid $\mathcal{E}_B$, Bob's Quantum Steering Ellipsoid (QSE) \cite{QSE2014}, which is characterised by an ellipsoid matrix ($Q_B$) and a centre ($\bm{c}_B$) which depend on the two-qubit state and are given by \cite{QSE2014}: 
\begin{align}
    \bm{c}_B
    =
    &
    \gamma_a^2
    (
    \bm{b}
    -
    T^T\bm{a}
    )
    ,\hspace{0.4cm}  
    \gamma_a
    =
    \frac{1}{\sqrt{1-a^2}}, \hspace{0.2cm} 
    a=|\bm{a}|, 
    \label{eq:QSEcentre}\\
    Q_B
    =&
    \gamma_a^2 
    (
    T^T-\bm{b}\bm{a}^T
    )(
    \mathds{1}
    +
    \gamma_a^2
    \bm{a} \,\bm{a}^T
    )(
    T-\bm{a} \bm{b}^T
    ),
    \label{eq:QSEmatrix}
\end{align} 
with $\bm{a}, \bm{b},T$ as in \eqref{eq:Rmatrix}. The square root of the eigenvalues of $Q_B$ correspond to the lengths of the semiaxes of the ellipsoid, whilst the eigenvectors correspond to the ellipsoid's orientation \cite{QSE2014}. Therefore, for a given two-qubit state $\rho$, Bob's QSE is specified by the pair $\mathcal{E}_B=\{Q_B, \bm{c}_B\}$. Similarly, we can calculate Alice's QSE $\mathcal{E}_A=\{Q_A, \bm{c}_A\}$ by switching $\bm{a} \leftrightarrow \bm{b}$ and replacing $T \rightarrow T^T$ in \eqref{eq:QSEcentre} and \eqref{eq:QSEmatrix} \cite{QSE2014}. We now address, in the same spirit of Ref. \cite{AM2014pra}, a conjecture that relates the QSE's centres and the violation of both the $\rm CHSH$-inequality and the $\rm F_3$-inequality. In words, this conjecture says that the maximal violation of the CHSH and $\rm F_3$ inequalities can be upper bounded by a function of the QSE centre magnitude \emph{alone}, without reference to the QSE matrix. This conjecture reads:

\begin{conjecture} (QSE conjecture)
For any two-qubit state $\rho \in D(\mathds{C}^2 \otimes \mathds{C}^2)$, its maximal violation of the $\rm CHSH$-inequality \eqref{eq:HC} and the $\rm F_3$-inequality \eqref{eq:CAC} are upper bounded by functions that depend on the magnitude of the QSE's centre as ${\rm B}(\rho) \leq f_{\rm CHSH}(c_B)$ and ${\rm F_3}(\rho)\leq f_{\rm F_3}(c_B)$, respectively. With $c_B=|\bm{c}_B|$.
\end{conjecture}

\begin{figure}[h!]
    \centering
    \includegraphics[scale=0.33]{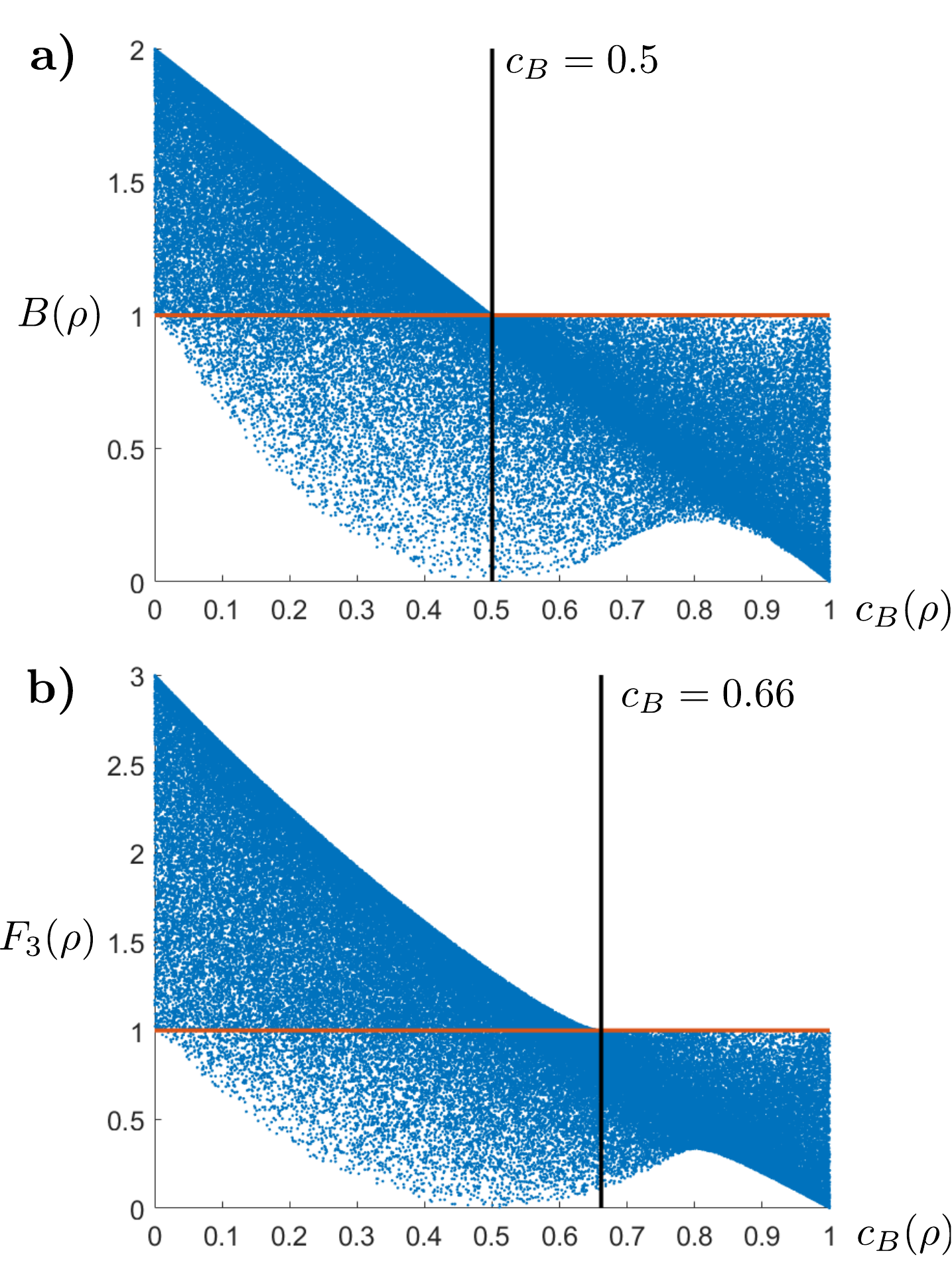}
    \vspace{-0.4cm}
    \caption{Conjecture on the relationship between two measures of quantum correlations versus the QSE's centre. {\bf a)} $\rm CHSH$-inequality violation \eqref{eq:HC} and {\bf b)} $\rm F_3$-inequality violation \eqref{eq:CAC} for $10^8$ randomly generated arbitrary two-qubit states against the magnitude of their respective QSE's centre ($c_B$) \eqref{eq:QSEcentre}. Similar plots can be obtained when analysing Alice's QSE centre $c_A$ (not shown). Horizontal orange lines at $1$ depict the classical bound for these inequalities. Vertical black lines depict the values $c_{\rm \, CHSH}=0.5$ and $c_{\rm \, F_3}= 0.66$ beyond of which states appear to no longer violate the respective inequality.}
    \label{fig:fig1}
\end{figure}

The part of the conjecture concerning the CHSH-inequality was first introduced in \cite{AM2014pra}, where the authors proposed the function $f_{\rm CHSH}(c_B)={\rm max} \{\sqrt{2(1-c_B)},1 \}$. This is an interesting upper bound which tells us that quantum states with QSE's centres $c_B>0.5$ ($f_{\rm CHSH}(c_B) = 1$) cannot violate the CHSH-inequality. Since QSEs are contained within the Bloch ball, this leads to the appealing geometric observation that states with QSE's centres ``close" ($c_B>0.5$) to the surface cannot violate the CHSH-inequality. In this work, we propose a similar conjecture for the violation of the ${\rm F}_3$-inequality. This conjecture for ${\rm F}_3$ has not been reported before, to the best of our knowledge, and it follows a similar behaviour to that of CHSH. Exploring the form of function $f_{\rm F_3}(c_B)$ is an interesting task which we leave for future research. In \autoref{fig:fig1} we show that the conjecture holds true for $10^8$ randomly generated arbitrary two-qubit states. In particular, we highlight that there seem to exist values $c_{\rm \, CHSH}=0.5$ and $c_{\rm \, F_3} = 0.66$, such that states with QSE centres' magnitudes greater than these values, can no longer lead to a violation of the $\rm CHSH$-inequality and the $\rm F_3$-inequality, respectively. Both bounds are extracted from the numerical results in \autoref{fig:fig1}. It is also worth mentioning that the authors in \cite{AM2014pra} proposed an additional conjecture for the measure of \emph{fully entangled fraction} which, although interesting in its own right, is omitted in this work since it does not display a similar impossibility region like $\rm CHSH$ or $\rm F_3$. In the next section we will exploit the existence of these impossibility regions for detecting locally inaccessible hidden quantum correlations. We now move on to introduce this phenomenon more formally.

\section{Hidden Quantum Correlations}

In this section we start by addressing the standard procedure for revealing hidden quantum correlations. We then move on to more restrictive versions, where only one-sided local filters are allowed, to then formalise what we mean by locally  \emph{accessible} and \emph{inaccessible} hidden quantum correlations. We then use the previous conjecture on QSEs to derive a sufficient criterion for detecting locally  inaccessible hidden quantum correlations. Specifically, locally  inaccessible hidden $\rm CHSH$-inequality violation and $\rm F_3$-inequality violation.

\subsection{Standard hidden quantum correlations}

As described in the previous section, consider experimentalists Alice and Bob sharing an arbitrary two-qubit state $\rho \in D(\mathds{C}^2 \otimes \mathds{C}^2)$ but now, before implementing a standard Bell-test, they implement a local filtering procedure as follows. Alice and Bob can each perform a local binary POVM measurement given by $E_W=\{E^0_W, E^1_W\}$, $W\in \{A,B\}$ where $E^1_W=\mathds{1}-E^0_W$, $E^0_W=f_W^{\dagger}f_W$, with $f_W$ satisfying the property $f_W^{\dagger} f_W \leq \mathds{1}$. This last property guarantees that the procedure can be performed as a valid measurement $E_W^1 \geq 0$. After the implementation of this measurement, the post-measured state is the unnormalised state $(f_A \otimes f_B)\rho (f_A\otimes f_B)^{\dagger}$ with probability $\tr[(f_A^\dagger f_A \otimes f_B^\dagger f_B)\rho]$. They then keep the post-measured state \emph{only} when they obtain this desired target state, discarding the system otherwise, and repeating until success. This procedure is also known as \emph{Stochastic Local Operations with Classical Communication (SLOCC)}, and the operators $f_A$ and $f_B$ as \emph{local filters} or SLOCC. This local filtering procedure then, can effectively be seen as transforming the initial state to a filtered state of the form:
\begin{align}
    \rho'
    =
    \frac{
    (f_A \otimes f_B)\rho (f_A \otimes f_B)^{\dagger}
    }{
    \tr
    \left[
    (f_A^{\dagger}f_A \otimes f_B^{\dagger}f_B) 
    \rho \right]
    },
    \label{eq:FS}
\end{align}
where $f_W \in {\rm GL}(2,\mathds{C})$, the group of invertible $2 \times 2$ complex matrices. This latter condition guarantees that the transformation does not destroy quantum correlations \cite{FV2001a}, $\rm CHSH$-nonlocality and $\rm F_3$-steering in particular \cite{FV2002prl}. Amongst all possible local filtering operations, there exists a particular one which we address here as the Kent-Linden-Massar (KLM0) SLOCC transformation \cite{LF5}, the KLM0-SLOCC from now on \footnote{We use the suffix $0$ in KLM0 in order to distinguish this acronym from the Knill-Laflamme-Milburn (KLM) protocol for linear optical quantum computation.}, which has the important property of transforming the state $\rho$ into its \emph{Bell-diagonal unique normal form} \cite{LF5}, which we address as $\rho^{\rm BD}_{\rm UNF}$ \cite{LF6, LF7}. It has been proven that the KLM0-SLOCC is the optimal local filtering transformation that \emph{simultaneously} maximises the quantum correlations of: concurrence \cite{LF6}, usefulness for teleportation \cite{LF7}, and the violation of the CHSH-inequality for Bell-nonlocality \cite{LF7}. The KLM0-SLOCC then effectively acts as $\rho \rightarrow \rho^{\rm BD}_{\rm UNF}$, which in terms of the R-picture \eqref{eq:Rmatrix} reads $R\rightarrow R^{\rm BD}_{\rm UNF}=\text{diag}(1,-\sqrt{\nu_1/\nu_0},-\sqrt{\nu_2/\nu_0},-\sqrt{\nu_3/\nu_0})$, with $\{\nu_{i=0,1,2,3}\}$ the eigenvalues of the operator $\eta R\eta R^T$ in decreasing order and $\eta=\text{diag}(1,-1,-1,-1)$ \cite{PG2015}. Hence, the quantum correlations of the Bell-diagonal unique normal form, which defines the {\it hidden quantum correlations} of the initial state $\rho$, are given by:
\begin{align}
    {\rm HB}^*(\rho)&
    \coloneqq
    {\rm B}\left (\rho^{\rm BD}_{\rm UNF}\right)
    =
    \sqrt{\frac{\nu_1+\nu_2}{\nu_0}},\label{eq:PGC}\\
    {\rm HF_3}^*(\rho)&
    \coloneqq 
    {\rm F_3}\left(\rho^{\rm BD}_{\rm UNF}\right)
    =
    \sqrt{\frac{\nu_1+\nu_2+\nu_3}{\nu_0}}. \label{eq:DSC}
\end{align}
As mentioned before, the KLM0-SLOCC is the optimal SLOCC for maximising the CHSH-inequality and therefore ${\rm HB}(\rho)\coloneqq\max_{\{f_A,f_B\}}{\rm B}(\rho')={\rm HB}^*(\rho)$, with $\rho'$ defined as in \eqref{eq:FS}. It is not known whether this is also the case for the $\rm F_3$-inequality, but we nevertheless have the inequality ${\rm HF_3}(\rho)\coloneqq\max_{\{f_A,f_B\}}{\rm F_3}(\rho')\geq {\rm HF_3}^*(\rho)$. 

\subsection{One-sided hidden quantum correlations}

We are now interested in restricting the standard locally filtered Bell-test scenario to the case when \emph{only one} of the parties is allowed or capable of implementing local filters. We define one-sided filtered states $\rho'_{\rm FW}$, $W\in \{A, B\}$ as:
{\small\begin{align*}
    \rho'_{\rm \,FA}
    =
    \frac{
    (f_A \otimes \mathds{1})
    \rho
    (f_A \otimes \mathds{1})^{\dagger}
    }{
    \tr
    \left[
    (f_A^{\dagger}f_A\otimes \mathds{1}) 
    \rho
    \right]
    },\hspace{0.2cm}
    \rho'_{\rm \,FB}
    =
    \frac{
    (\mathds{1}\otimes f_B)\rho (\mathds{1} \otimes f_B)^{\dagger}
    }{
    \tr
    \left[
    (\mathds{1} \otimes f_B^{\dagger}f_B ) 
    \rho
    \right]
    }.
\end{align*}}
These two locally filtered states can alternatively be seen as imposing the condition $f_B=\mathds{1}$ or $f_A=\mathds{1}$ in the standard definition \eqref{eq:FS}, respectively. We can now define two measures for locally  accessible hidden $\rm CHSH$-inequality violation, and two measures for locally  accessible $\rm F_3$-inequality violation as follows:
\begin{align}
    {\rm HB_W}(\rho)
    &:=
    \underset{\{f_W\}}{{\rm max}} \, {\rm B}(\rho'_{\rm \, FW}), \hspace{0.3cm} W \in \{A, B\},
    \label{eq:HBA}\\
    {\rm HF_{3W}}(\rho)
    &:=
    \underset{\{f_W\}}{{\rm max}} \, {\rm F_3}(\rho'_{\rm \, FW}), \hspace{0.3cm} W \in \{A, B\}.
\label{eq:HF3A}
\end{align}
These measures define the amount of hidden correlations that each party (either Alice or Bob) can extract by working unilaterally while the other party does nothing. It follows from these definitions that we have the inequalities ${\rm B}(\rho)\leq {\rm HB_W}(\rho) \leq {\rm HB}(\rho)$, and similarly for the $\rm F_3$-inequality. It would be desirable to have analytical expressions for these asymmetric measures, as it is the case for their symmetric counterparts. We are now interested in the scenarios where it is possible to reveal quantum correlations when they were not initially present. In \autoref{table:table1} we define five fine-grained hidden quantum correlation phenomena. 

\begin{widetext}
	\begin{center}
		\begin{table}[h!]
		\begin{tabular}{|c||c|c|c|c|c||c|}
				\hline
				\textbf{Case}&
				\textbf{Definition:} $\rho$ displays...&
				${\rm B}(\rho)$        \eqref{eq:HC} & 
				${\rm HB_A}(\rho)$  \eqref{eq:HBA}&  
				${\rm HB_B}(\rho)$  \eqref{eq:HBA}  &
				${\rm HB}(\rho)$     \eqref{eq:PGC} &
				\textbf{Reference}\\ 
				\hline
				\hline
				1&
				Hidden CHSH&
				$\leq 1$ & 
				-&  
				-&
				$> 1$ &  
				\cite{SP1995, Gisin1996}\\ 
				\hline
				2&
				Maximal hidden CHSH&
				$\leq 1$ & 
				-&  
				-&
				$ = 2$ &
				\cite{DSR2020b, erasure1, erasure2}\\ 
				\hline
				3&
				B-inaccessible hidden CHSH&
				$\leq 1$ & 
				$ > 1$ &  
				$\leq 1$ &
				$> 1$ &  
				This work
                    (\autoref{sec:unilateral_inaccesible})\\
				\hline
				4&
				AB-inaccessible hidden CHSH &
				$\leq 1$ & 
				$\leq 1$ &  
				$\leq 1$ &
				$ > 1$ &  
				This work
                    (\autoref{sc:locally_ab})\\
				\hline
				5&
				AB-inaccessible maximal hidden CHSH&
				$\leq 1$ & 
				$\leq 1$ &  
				$\leq 1$ &
				$ = 2$ &  
				This work 
                    (\autoref{sc:locally_max})\\ 
				\hline
			\end{tabular}
			\caption{Summary of definitions and results on fine-grained hidden quantum correlations regarding the violation of the CHSH-inequality for Bell-nonlocality. We can similarly define these cases for the $\rm F_3$-inequality violation for EPR-steering. 
            }
		\label{table:table1}
		\end{table}
	\end{center} \vspace{-1cm}
\end{widetext} 
We now describe the cases in \autoref{table:table1}. The first case addresses standard hidden CHSH-inequality violation, which was first introduced by Popescu in \cite{SP1995} and specialised to two-qubit states by Gisin in \cite{Gisin1996}. The second case deals with \emph{maximal} hidden CHSH-inequality violation.  An example of this phenomenon was shown to be present in a qutrit-qubit state that goes by the name of the \emph{erasure state} \cite{erasure1, erasure2}. A case for two-qubit states was considered in \cite{DSR2020b}, as states coming from the dynamics of open quantum systems. The third case specifies \emph{locally  B-inaccessible} hidden CHSH-inequality violation, which encapsulates the idea of the quantum state not allowing Bob to enhance its correlations, no matter what local filter he is using, so that the cooperation of Alice is indispensable. Here we also naturally define locally  \emph{A-accessible} hidden $\rm CHSH$-inequality violation, meaning that Alice alone can extract some amount of correlations. The fourth case considers the stronger notion of a state whose hidden quantum correlations are locally  inaccessible from \emph{both} sides. This encapsulates the idea of the quantum state not allowing Alice and Bob to act individually, but forcing them to cooperate, and hence the wording \emph{locally AB-inaccessible} hidden $\rm CHSH$-inequality violation. We emphasise here that the difference between standard hidden quantum correlations (for which we have ${\rm HB}(\rho) > 1$) and AB-inaccessible hidden quantum correlations (for which we also have ${\rm HB}(\rho) > 1$) is that, in the latter, we can additionally guarantee that ${\rm HB_A}(\rho)\leq 1$ and ${\rm HB_B}(\rho)\leq 1$. Finally, the fifth case considers an extreme case in which, the correlations are locally  inaccessible and yet, they can still be revealed to be the maximum amount allowed by quantum theory.

These one-sided versions here introduced can be found useful when considering semi-device independent protocols, as it has been the case for EPR-steering \cite{QS2017, review_steering2}. Unlike the standard hidden correlations, for which we have the closed formulas in \eqref{eq:PGC} and \eqref{eq:DSC}, there are currently no closed formulas for the one-sided versions defined in \eqref{eq:HBA} and \eqref{eq:HF3A}. In this work we start exploring these alternative one-sided measures. The first step we take is to address whether there actually exist states with the properties depicted in these three definitions. This is because, a priori, it might well be the case that the hidden correlations of all states are actually already always locally accessible (by both parties) and therefore, the previous definitions are unnecessary. The main challenge in tackling these questions is that, currently, there are no efficient tools for guaranteeing that the hidden correlations of a state are locally inaccessible or, explicitly, that ${\rm HB_A}(\rho)\leq 1$ or that ${\rm HF_{3A}}(\rho) \leq 1$. In this work we take a first step in this direction, by providing a sufficient criterion for guaranteeing that a state has \emph{locally  $W$-inaccessible hidden quantum correlations} with $W \in \{A,B\}$, so that when considered together, it also allows us to explore locally AB-inaccessible hidden quantum correlations. 

\subsection{Sufficient criterion for locally  $W$-inaccessible hidden quantum correlations $W\in \{A, B\}$}

We now provide a sufficient criterion for guaranteeing that a state possesses locally  W-inaccessible hidden quantum correlations. We first need the following lemma about quantum steering ellipsoids \cite{QSE2014}.
\begin{lemma} 
Consider a two-qubit state $\rho \in D(\mathds{C}^2 \otimes \mathds{C}^2)$ with associated QSEs given by $\mathcal{E}_W=\{\bm{c}_W, Q_W\}$, $W\in\{A, B\}$. Consider also that the state is locally filtered to a state $\rho'$ with QSEs given by $\mathcal{E}'_W=\{\bm{c}_W', Q'_W\}$, $W\in \{A,B\}$. If we consider local filters of the form $\mathds{1}\otimes f_B$, then $\mathcal{E}'_A = \mathcal{E}_A$. Conversely, if we consider local filters of the form $f_A \otimes \mathds{1}$, then $\mathcal{E}'_B = \mathcal{E}_B$. 
\end{lemma}
The proof of this lemma can be found in \cite{QSE2014}. We now use this lemma to establish our main result.
\begin{result} \label{r:r1}
Consider a two-qubit state $\rho \in D(\mathds{C}^2\otimes \mathds{C}^2)$. If Alice's (Bob's) QSE centre magnitude satisfies $c_A>c_{\rm \,CHSH}$ ($c_B>c_{\rm \, CHSH}$) with $c_{\rm \, CHSH}= 0.5$, then, modulo the QSE conjecture being true, it follows that $\,{\rm HB_B}(\rho) \leq 1$ (${\rm HB_A}(\rho)\leq 1$). This means that Bob (Alice) cannot locally  achieve any hidden $\rm CHSH$-inequality violation.
\end{result}

\begin{proof}
Consider a two-qubit state $\rho \in D(\mathds{C}^2\otimes \mathds{C}^2)$ with Alice's QSE centre satisfying $c_A>c_{\rm CHSH}=0.5$. Then, because of the QSE conjecture, the state cannot violate the CHSH-inequality. Moreover, if we consider local filters of the form $\mathds{1} \otimes f_B$ and the associated filtered state $\rho^{'}_{\rm FB}$, the previous Lemma guarantees that Alice's QSE $\mathcal{E}_A$ remains unchanged as $\mathcal{E}'_A = \mathcal{E}_A$. In particular, the magnitude of the QSE's centre remains unchanged and, therefore, again by the QSE conjecture, the filtered state $\rho^{'}_{\rm FB}$ cannot violate the $\rm CHSH$-inequality, meaning that ${\rm B}(\rho^{'}_{\rm FB})\leq 1$. Since this holds for any $f_B$, it therefore follows that ${\rm HB_B}(\rho) = \max_{f_B}{\rm B}(\rho^{'}_{\rm FB})\leq 1$, thus completing the proof. The same argument holds with Bob's QSE centre satisfying $c_B>c_{\rm CHSH}=0.5$, but now for ${\rm HB_A}(\rho)$.
\end{proof}
Taking these two criteria together, it allows us to look for locally AB-inaccessible hidden $\rm CHSH$-inequality violation by calculating the QSE's centre magnitudes $c_A$ and $c_B$. The F3 inequality also displays an impossibility region, and so we have an analogous result.
\begin{result} \label{r:r2}
Consider a two-qubit state $\rho \in D(\mathds{C}^2\otimes \mathds{C}^2)$. If the QSE centre's magnitude satisfies $c_A>c_{\rm \,F_3}$ ($c_B>c_{\rm \, F_3}$) with $c_{\rm \, F_3}= 0.66$, then, modulo the QSE conjecture being true, we have that ${\rm HF_{3B}}(\rho) \leq 1$ ($\, {\rm HF_{3A}} (\rho) \leq 1$). This means that Bob (Alice) cannot locally  access any hidden $\rm \rm F_3$-inequality violation.
\end{result}

With these sufficient criteria in place, it becomes a straightforward exercise to look for both locally inaccessible hidden $\rm CHSH$-inequality and $\rm F_3$-inequality violation, since it all boils down to calculate the QSE's centres, which are explicitly given as per \eqref{eq:QSEcentre}, and comparing these values to $c_{\rm CHSH}=0.5$ and $c_{\rm F_3}=0.66$, respectively. We now proceed to use these sufficient criteria to explore locally W-inaccessible, $W\in \{A, B\}$, and locally AB-inaccessible hidden quantum correlations of two-qubit states, showing that these sufficient criteria are enough detect the existence of states displaying such properties.

\section{Examples: Locally inaccessible hidden correlations}

We now consider specific two-qubit states and calculate the properties of entanglement by means of the Positive Partial Transpose (PPT) criterion \cite{PPT1996}, $\rm CHSH$-inequality violation \eqref{eq:HC}, $\rm F_3$-inequality violation \eqref{eq:CAC}, hidden $\rm CHSH$-inequality violation \eqref{eq:PGC}, hidden $\rm F_3$-inequality violation \eqref{eq:DSC}, unsteerability \cite{PS2016, NB2016}, and the regions for which the QSEs' centres \eqref{eq:QSEcentre} satisfy the conditions $c_A,c_B>c_{\rm CHSH}= 0.5$ and $c_A,c_B>c_{\rm F_3}= 0.66$ so to guarantee the locally  $W$-inaccessibility $W \in \{A,B\}$ of either CHSH-nonlocality or $\rm F_3$-steering, respectively, as per \cref{r:r1} and \cref{r:r2}. We address three examples. First, an asymmetric case which displays A-accessible, B-inaccessible hidden quantum correlations. Second, a symmetric case which displays locally AB-inaccessible hidden quantum correlations. Third and finally, an extreme case of states that display locally AB-inaccessible \emph{maximal} hidden quantum correlations. 

\subsection{Locally  A-accessible, B-inaccessible \\
hidden quantum correlations}
\label{sec:unilateral_inaccesible}

\begin{figure}[h!]
    \centering
    \includegraphics[scale=0.55]{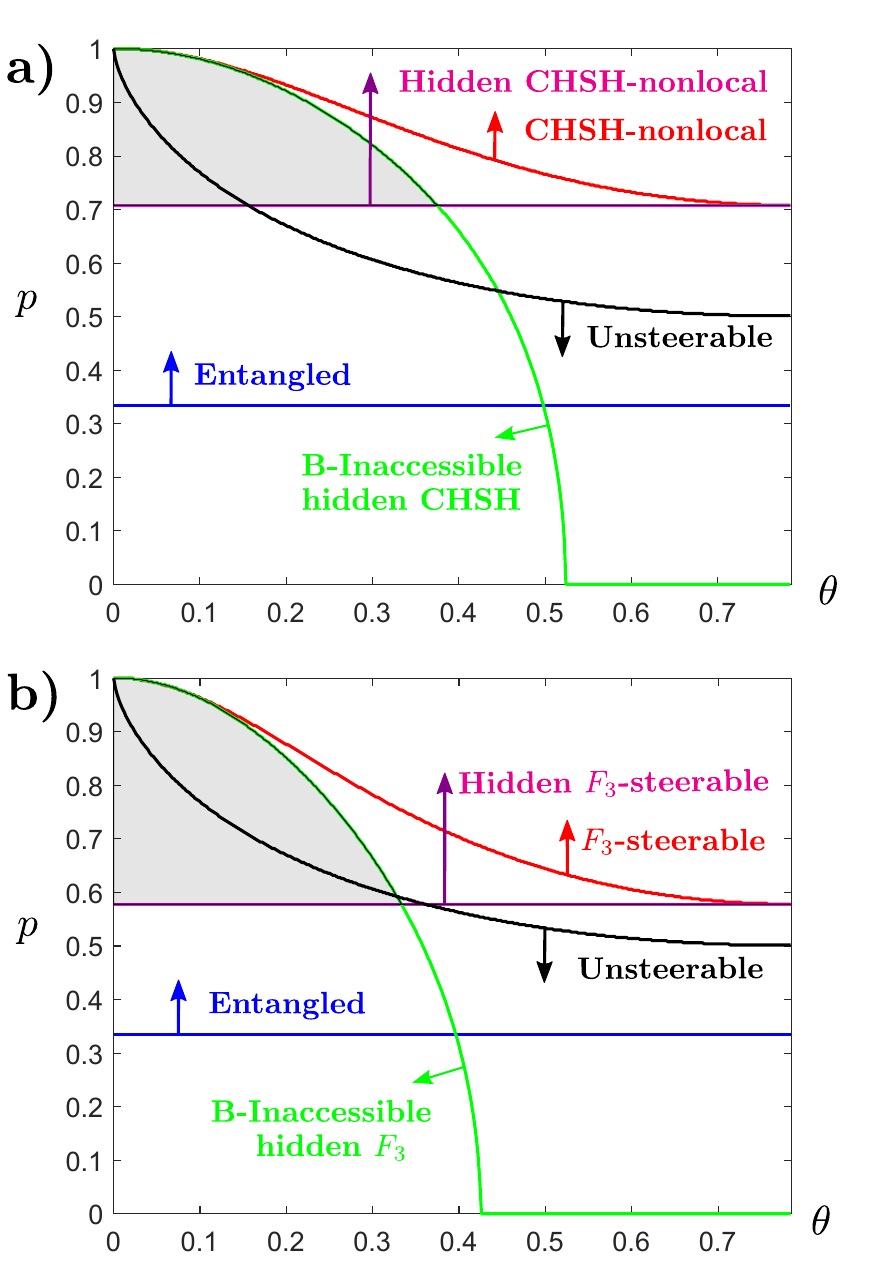}
    \vspace{-0.5cm}
    \caption{
    Quantum correlation measures of partially entangled states with coloured noise \eqref{eq:MS}. In both subfigures, entanglement in blue is calculated by means of the PPT criterion, and unsteerability in black by means of the sufficient criterion from \cite{criterion_UN}. \textbf{a)} CHSH-inequality related correlation measures: in red the $\rm CHSH$-inequality violation \eqref{eq:HC}, in magenta the hidden $\rm CHSH$-inequality violation \eqref{eq:PGC}, in green the QSE centre condition $c_A>c_{\rm CHSH}= 0.5$. The shaded region shows states with locally  B-inaccessible (yet locally  A-accessible) hidden CHSH-inequality violation as per \cref{r:r1} (see case 3 of \cref{table:table1}). \textbf{b)} $\rm F_3$-inequality related correlation measures: in red the $\rm F_3$-inequality violation \eqref{eq:CAC}, in magenta the hidden $\rm F_3$-inequality violation \eqref{eq:DSC}, in green the QSE centre condition $c_A>c_{\rm F_3}= 0.66$. The shaded region depicts states with locally B-inaccessible $\rm F_3$-inequality violation as per \cref{r:r2}.
    } 
    \label{fig:fig2}
\end{figure}

We show evidence the existence of this phenomenon with a particular family of states which is usually called partially entangled states with (asymmetric) coloured noise:
\begin{align}
    \rho_{\rm M}(\theta, p)=p\phi^+(\theta)+(1-p)\rho_A(\theta)\otimes \mathds{1},
    \label{eq:MS}
\end{align}
where $\phi^+(\theta)=\ketbra{\phi^+(\theta)}{\phi^+(\theta)}$, $\rho_A(\theta)=\tr_{B}[\phi^+(\theta)]$, $0\leq \theta \leq \pi/4$ and $0 \leq p\leq 1$, $\ket{\phi^+(\theta)}=\cos \theta \ket{00}+\sin \theta \ket{11}$. In \autoref{fig:fig2} \textbf{a)} we address $\rm CHSH$-inequality related correlations and, particularly, we distinguish regions for $c_A>c_{\rm CHSH}=0.5$, that is, A-accessible, B-inaccessible hidden $\rm CHSH$-inequality violation. In \autoref{fig:fig2} \textbf{b)} we address $\rm F_3$-inequality related correlations with regions for locally A-accessible, B-inaccessible hidden $F_3$-inequality violation ($c_A>c_{F_3}=0.66$). In both subfigures the shaded region refers to states displaying A-accessible, B-inaccessible hidden CHSH-inequality and $F_3$-inequality violation, respectively. The KLM0-SLOCC for these states can explicitly be written as:
\begin{align*}
    f_A(\theta, p)= \sin \theta \begin{pmatrix} \frac{1}{\cos \theta} &0 \\ 0 & \frac{1}{\sin \theta} \end{pmatrix},
    \hspace{0.5cm}
    f_B(\theta, p)= \begin{pmatrix} 1 &0 \\ 0 & 1 \end{pmatrix}.
\end{align*}
We have that these states are locally A-accessible as the optimal local filters do not require that Bob acts on his part, in agreement with \autoref{fig:fig2}. In this example, it is worth pointing out that the states below the black curve (both subfigures)  admit a LHS model and so, beyond hidden CHSH-inequality violation, we more generally have instances of hidden Bell-nonlocality.

\subsection{Locally AB-inaccessible
hidden quantum correlations}
\label{sc:locally_ab}

\begin{figure}[h!]
    \centering
    \includegraphics[scale=0.55]{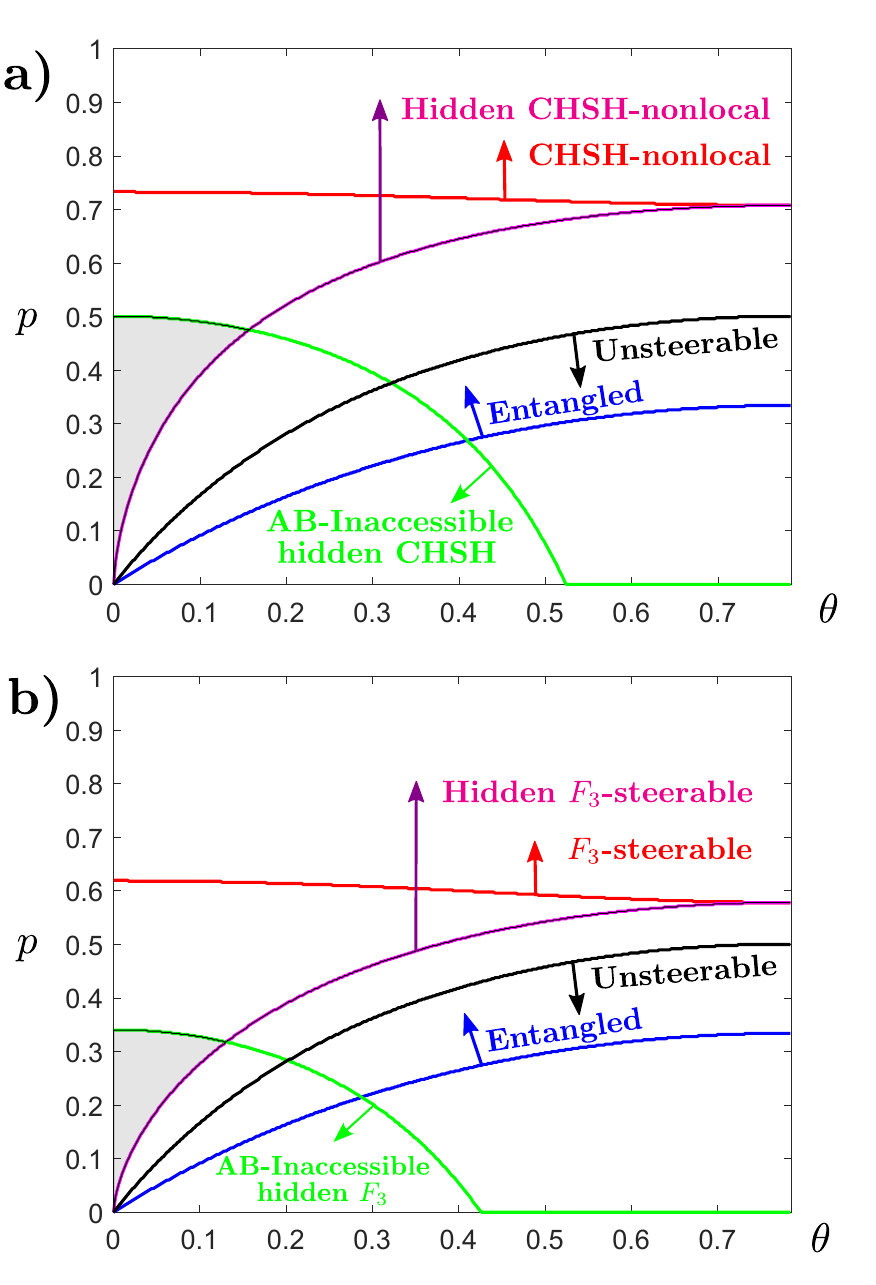}
    \vspace{-0.5cm}
    \caption{
    Quantum correlation measures of partially entangled states with symmetric coloured noise \eqref{eq:MMS}. In both subfigures,
    entanglement, unsteerability, $\rm CHSH$-inequality, $\rm F_3$-inequality as well as hidden inequalities are labelled as in \cref{fig:fig2}. \textbf{a)} Green curve indicates the QSE centre condition $c_A, c_B>c_{\rm CHSH}= 0.5$. In this case, the shaded region corresponds to states with locally AB-inaccessible hidden $\rm CHSH$-inequality violation as per \cref{r:r1}. This means that neither Alice nor Bob can extract unilaterally hidden correlations but they are required to cooperate (case 4 in \cref{table:table1}). \textbf{a)} Green curve indicates the QSE centre condition $c_A, c_B>c_{\rm CHSH}= 0.66$ such that shaded region corresponds to states with locally AB-inaccessible hidden $\rm F_3$-inequality violation as per \cref{r:r2}.
    }
    \label{fig:fig3}
\end{figure}

We now modify the previous states to be partially entangled states with symmetric coloured noise as:
\begin{align}
    \rho_{\rm MM} 
    (\theta,p) 
    =p\,
    \phi^+(\theta)
    +
    (1-p)
    \rho_A(\theta)
    \otimes
    \rho_B(\theta),
    \label{eq:MMS}
\end{align}
with $\rho_A(\theta)=\tr_{B}[\phi^+(\theta)]$, $\rho_B(\theta)=\tr_{A}[\phi^+(\theta)]$, and $0 \leq p\leq 1$. In \autoref{fig:fig3} \textbf{a)} we address CHSH-inequality related correlations whilst subfigure \textbf{b)} deals with the $\rm F_3$-inequality related correlations.  Unlike the previous case, we now have regions for locally AB-inaccessible CHSH-inequality and $\rm F_3$-inequality violation, as opposed to only B-inaccessible hidden quantum correlations. Furthermore, in the shaded regions we have B-inaccessible hidden CHSH-inequality and $F_3$-inequality violation, respectively.

\subsection{Locally AB-inaccessible
maximal hidden quantum correlations}
\label{sc:locally_max}

\begin{figure}
    \centering
    \includegraphics[scale=0.8]{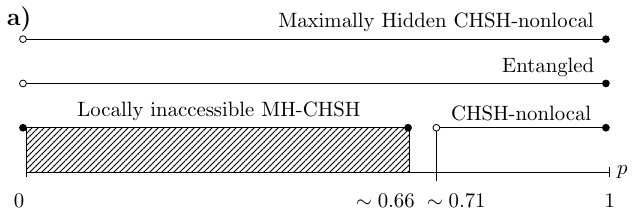}
    \includegraphics[scale=0.8]{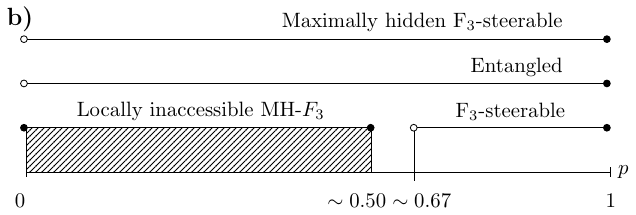}
    \vspace{-0.8cm}
    \caption{
    Quantum correlations of quasi-distillable states \eqref{eq:QDS}. In both subfigures entanglement is calculated by means of the PPT criterion. \textbf{a)} $\rm CHSH$-inequality related measures: $\rm CHSH$-inequality violation \eqref{eq:HC}, hidden $\rm CHSH$-inequality violation \eqref{eq:PGC}, locally AB-inaccessible \emph{maximal} hidden $\rm CHSH$-inequality violation (shaded region) as per \cref{r:r1}. \textbf{b)} $\rm F_3$-inequality related measures: $\rm F_3$-inequality violation \eqref{eq:CAC}, hidden $\rm F_3$-inequality violation \eqref{eq:DSC}, locally AB-inaccessible maximal hidden $\rm F_3$-inequality violation (shaded region) as per \cref{r:r2}. 
    }
    \label{fig:fig4}
\end{figure}

We now address the so-called quasi-distillable states \cite{FV2002prl} which can be parametrised by $0 \leq p\leq 1$ as:
\begin{align}
    \rho_{\rm QD}
    (p)
    =
    p\,
    \ketbra{\Psi^-}{\Psi^-}
    +
    (1-p)
    \ketbra{00}{00}
    ,
    \label{eq:QDS}
\end{align}
with $\ket{\Psi^-}=(1/\sqrt{2})(\ket{01}-\ket{10})$. These states belong to the SLOCC-orbit of the singlet, so in this sense they can be distilled \cite{FV2002prl}. This procedure however has a success probability that decreases with $p \rightarrow 0$, hence the name quasi-distillable. In \autoref{fig:fig4} \textbf{a)} we address CHSH-inequality related correlations and, in particular, in the shaded region ($0<p<0.66$) they evidence locally AB-inaccessible \emph{maximal} hidden CHSH-inequality violation. In \autoref{fig:fig4} \textbf{b)} we address  $\rm F_3$-inequality related correlations with the shaded region ($0<p<0.50$) evidencing locally AB-inaccessible \emph{maximal} hidden $F_3$-inequality violation. We emphasise that the states in the shaded region go from having weak entanglement, CHSH-inequality, and $F_3$-inequality violation, to being the singlet state and therefore, having the maximum amount of these correlations that is allowed by quantum theory. Overall, these simple families of states show that this phenomenon isn't hard to find, and so we believe it is in fact an ubiquitous aspect of the correlations allowed by quantum theory.

\section{conclusions}

In this work we prove, modulo a conjecture on quantum steering ellipsoids being true, the existence of the phenomenon of \emph{locally inaccessible hidden quantum correlations}. Particularly, locally inaccessible hidden {\rm CHSH}-inequality violation for Bell-nonlocality and $\rm F_3$-inequality violation for EPR-steering.  The case of simultaneous A-inaccessible and B-inaccessible hidden quantum correlations can alternatively be regarded as a type of ``super" hidden nonlocality, in the sense that it is a type of nonlocality that is actually \emph{more} hidden than its standard counterpart, this, since it explicitly requires the active intervention of \emph{both} parties, as opposed to potentially only one of them. Moreover, we report on a stronger version of this phenomenon in the form of locally inaccessible \emph{maximal} hidden quantum correlations, meaning that the local filters are revealing the maximal amount of correlations allowed by quantum theory. The relatively simple families of states that display these phenomena we provided here were not difficult to find, and so this leads us to believe that this is in actuality a generic feature of the type of correlations allowed by quantum theory. 

We believe that the results found in this work open up several questions for future research. First, although the QSE's conjectures in question are supported by numerical results, they however lack an analytical proof. It would therefore be desirable to have analytical proofs for these QSE's conjectures, so in order to completely guarantee the existence of this phenomenon. Second, it would also be interesting to derive closed formulas for these locally accessible hidden quantum correlations measures, as it has been done for their standard hidden counterparts. Third, the idea of quantum correlations being revealed by filters on only one side of the experiment can naturally be extended to other setups like: higher dimensions, multipartite scenarios, as well as to other measures of correlations like: entanglement, quantum obesity, quantum discord and so on. We remark however, that the cases for entanglement and quantum obesity for instance, do \emph{not} display a \emph{hidden phenomenon}, but a rather a \emph{extractable} counterpart \cite{LF5}. This, in the sense that local filters cannot take separable states into entangled states, but it can nonetheless take entangled states into states with larger amount of entanglement. It would nonetheless be desirable to have closed formulas for these cases as well. Fourth, in addition to the hidden CHSH-inequality (${\rm F_3}$-inequality) violation, one can define stronger versions of these phenomena in the form of hidden Bell-nonlocality (EPR-steering), by guaranteeing that the pre-filtered state allows LHV and LHS models. This latter can be explored for instance, for two-qubit states, by means of the sufficient unsteerability criterion derived in \cite{criterion_UN} and, for general states, by means of the numerical codes from \cite{PS2016, NB2016, NB2018}. Fifth and finally, from a practical point of view, the results found in this work could also find application in semi-device independent information-processing protocols, as it has already proven to be the case for EPR-steering. 

\emph{Acknowledgements.---}A.F.D. thanks Sabine Wollmann, David Payne, and Joel Tasker for discussions on quantum steering ellipsoids. A.F.D. and C.E.S. thank John H. Reina for discussions on quantum correlations. A.F.D. acknowledges support from the International Research Unit of Quantum Information, Kyoto University, the Center for Gravitational Physics and Quantum Information (CGPQI), and COLCIENCIAS 756-2016. C.E.S. acknowledges support from University of C\'ordoba (Grants FCB-12-23 and FCB-09-22). Part of this work was carried out whilst A.F.D. was a PhD student in the Quantum Engineering Centre for Doctoral Training (QE-CDT). P.S. acknowledges support from a Royal Society URF (NFQI). P.S. is a CIFAR Azrieli Global Scholar in the Quantum Information Science Programme.  

\bibliographystyle{apsrev4-1}
\bibliography{bibliography.bib}

\end{document}